\documentclass{article}
\usepackage{amssymb}
\usepackage{graphicx}
\usepackage{cite}
\begin{document}

\title{On the boundary conditions in deformed quantum mechanics with
minimal length uncertainty}
\author{Pouria Pedram\thanks{Email: p.pedram@srbiau.ac.ir}
\\  {\small Department of Physics, Science and Research Branch, Islamic
        Azad University, Tehran, Iran}}

\maketitle \baselineskip 24pt

\begin{abstract}
We find the coordinate space wave functions, maximal localization
states, and quasiposition wave functions in a GUP framework that
implies a minimal length uncertainty using a formally self-adjoint
representation. We show that how the boundary conditions in
quasiposition space can be exactly determined from the boundary
conditions in coordinate space.

\vspace{5mm} {\it PACS numbers:} 04.60.-m

\vspace{.5cm} {\it Keywords}: Generalized Uncertainty Principle;
Minimal length; Boundary conditions.
\end{abstract}
\maketitle

\section{Introduction}
The existence of a minimal length uncertainty proportional to the
Planck length $\ell_P\sim10^{-35}m$ is one of the common predictions
of various candidates of quantum gravity. This idea in the context
of the Generalized (Gravitational) Uncertainty Principle (GUP) has
attracted much attention in recent years and many papers have been
appeared in the literature to address the effects of this minimal
length on various quantum mechanical systems
\cite{felder,1,2,3,4,5,6,7,8,9,10,11,12,13}.

%**************************
It is pointed out by Mead that gravity amplifies the Heisenberg's
measurement uncertainty which makes it impossible to measure
distances more accurate that Planck's length \cite{Mead}. In fact,
since the increase of the energies to probe small distances
considerably disturbs the spacetime structure because of the
gravitational effects, the spatial uncertainty eventually increases
at energy scales as large as the Planck scale. This minimal length
can be considered as a fundamental property of quantum spacetime, a
natural UV-regulator, and a solution for the transplanckian problem.
Since the string theory with large or warped extra dimensions can
lower the Planck scale into the TeV range, this fundamental length
scale also moved into the reach of the Large Hadron Collider.

The thought experiments that support the minimal length proposal
include the Heisenberg microscope with Newtonian gravity and its
relativistic counterpart \cite{Mead}, limit to distance measurements
\cite{Wigner}, limit to clock synchronization, and limit to the
measurement of the black-hole–horizon \cite{Magg}. Moreover,
different approaches to quantum gravity such as string theory, loop
quantum gravity and loop quantum cosmology, quantized conformal
fluctuations \cite{Padmanabhan1,Padmanabhan2}, asymptotically safe
gravity \cite{hooft}, and non-commutative geometry all indicate a
fundamental limit to the resolution of structure.

%************************

Based on the Heisenberg's microscope and taking into account both
the normal and the gravitational uncertainties one finds
\cite{Adler}
\begin{eqnarray}
\Delta X \gtrsim \frac{1}{2\Delta P} +G\Delta P,
\end{eqnarray}
where $G=\ell_P^2$ is the gravitational constant. As Adler and
Santiago observed this GUP is invariant under $\ell_P \Delta
P\leftrightarrow \frac{1}{\ell_P \Delta P}$ and therefore has a
momentum inversion symmetry. Because of the universality of the
gravity, this correction modifies all Hamiltonians for the quantum
systems near the Planck scale.

Recently, an experimental scheme is suggested by Pikovski \emph{et
al.} to test the presence of the minimal length scale in the context
of quantum optics \cite{laser4}. They used quantum optical control
and optical interferometric techniques for direct measurement of the
canonical commutator deformations of a massive object. This
experiment does not need the Planck-scale accuracy of position
measurement and can be reached by the current technology. Some
attempts have been also made to test possible quantum gravitational
phenomena using astronomical observations \cite{laser3,laser3-2}.

In this paper, we consider a GUP that implies a minimal length
uncertainty proportional to the Planck length. We find the exact
coordinate space wave functions and quasiposition space wave
functions using a formally self-adjoint representation. We first
obtain the eigenfunctions of the position operator and the maximal
localization states. Then we discuss how the boundary conditions can
be imposed consistently in both coordinate space and quasiposition
space.

\section{The Generalized Uncertainty Principle}
Consider the following one-dimensional deformed commutation relation
\cite{12}
\begin{eqnarray}\label{gupc}
[X,P]=i\hbar(1+\beta P^2),
\end{eqnarray}
where for $\beta=0$ we recover the well-known commutation relation
in ordinary quantum mechanics and $\Delta X \Delta P \geq
\frac{\hbar}{2}\left(1+ \beta(\Delta P)^2 \right)$. Since $\Delta X$
cannot be made arbitrarily small, the absolutely smallest
uncertainty in positions for this GUP is $(\Delta
X)_{min}=\hbar\sqrt{\beta}$. To proceed further, consider the
following representation \cite{10}
\begin{eqnarray}\label{x0p01}
X &=& x,\\ P &=&
\frac{\tan\left(\sqrt{\beta}p\right)}{\sqrt{\beta}},\label{x0p02}
\end{eqnarray}
which exactly satisfies Eq.~(\ref{gupc}). This representation is
\emph{formally} self-adjoint subject to the inner product
\begin{eqnarray}\label{inner}
\langle\psi|\phi\rangle=\int_{-\frac{\pi}{2\sqrt{\beta}}}^{+\frac{\pi}{2\sqrt{\beta}}}\mathrm{d}p\,\psi^{*}(p)\phi(p),
\end{eqnarray}
and preserves the ordinary nature of the position operator.

The operator $A$ with dense domain ${\cal D}(A)$ is self-adjoint if
${\cal D}(A) ={\cal D}(A^{\dagger})$ and $A=A^{\dagger}$. However,
for the position operator $X$ in the momentum space we have
\begin{eqnarray}
&&\int_{-\frac{\pi}{2\sqrt{\beta}}}^{+\frac{\pi}{2\sqrt{\beta}}}\mathrm{d}p\,\psi^{*}(p)\left(i\hbar\frac{\partial}{\partial
p}\right)\phi(p)=
\int_{-\frac{\pi}{2\sqrt{\beta}}}^{+\frac{\pi}{2\sqrt{\beta}}}\mathrm{d}p\,\left(i\hbar\frac{\partial\psi(p)}{\partial
p}\right)^{*}\phi(p)+i\hbar\,\psi^*(p)\phi(p)\Bigg|_{p=+\frac{\pi}{2\sqrt{\beta}}}\hspace{-.5cm}-i\hbar\,\psi^*(p)\phi(p)\Bigg|_{p=-\frac{\pi}{2\sqrt{\beta}}},
\end{eqnarray}
where $\phi(p)$ vanishes at $p=\pm\frac{\pi}{2\sqrt{\beta}}$ and
$\psi^*(p)$ takes arbitrary values at the boundaries. Indeed, the
adjoint of the position operator
$X^{\dagger}=i\hbar\partial/\partial p$ has the same form but it
acts on a different space of functions
\begin{eqnarray}
{\cal D}(X)&=&\bigg\{\phi,\phi'\in{\cal
L}^2\left(\frac{-\pi}{2\sqrt{\beta}},\frac{+\pi}{2\sqrt{\beta}}\right)\,;
\phi\left(\frac{+\pi}{2\sqrt{\beta}}\right)=\phi\left(\frac{-\pi}{2\sqrt{\beta}}\right)=0\bigg\},\\
{\cal D}(X^{\dagger})&=&\bigg\{\psi,\psi'\in{\cal
L}^2\left(\frac{-\pi}{2\sqrt{\beta}},\frac{+\pi}{2\sqrt{\beta}}\right);\mbox{no
other restriction on }\psi\bigg\}.
\end{eqnarray}
Therefore, $X$ is merely symmetric ($X=X^\dagger$), but it is not a
true self-adjoint operator. On the other hand, the momentum operator
is a self-adjoint operator which can be shown using the von
Neumann's theorem \cite{10}, i.e., $P=P^\dagger$ and
\begin{eqnarray}
{\cal D}(P)={\cal D}(P^{\dagger})=\left\{\phi\in{\cal
D}_{max}\left(\mathbb{R}\right)\right\},
\end{eqnarray}
where ${\cal D}_{max}$ denotes the maximal domain on which $P$ has a
well defined action
\begin{eqnarray}
{\cal D}_{max}(P)=\left\{\phi\in{\cal L}^2(\mathbb{R}):P\phi\in{\cal
L}^2(\mathbb{R})\right\}.
\end{eqnarray}

In this representation, the completeness relation and scalar product
can be written as
\begin{eqnarray}\label{comp1}
\langle p'|p\rangle= \delta(p-p'),\\
\int_{-\frac{\pi}{2\sqrt{\beta}}}^{+\frac{\pi}{2\sqrt{\beta}}}\mathrm{d}
p\, |p\rangle\langle p|=1.\label{comp2}
\end{eqnarray}
Also the eigenfunctions of the position operator in momentum space
are given by the solutions of the eigenvalue equation
\begin{eqnarray}
X\,u_x(p)=x\,u_x(p),
\end{eqnarray}
where $u_x(p)=\langle p|x\rangle$. The normalized solution is
\begin{eqnarray}
u_x(p)=\sqrt{\frac{\sqrt{\beta}}{\pi}}\exp\left({-i\frac{ p}{\hbar}}
x\right).
\end{eqnarray}
Note that the physical meaning of the present eigenstates is
different from the ones provided in Refs.~\cite{10,12}. Now using
Eq.~(\ref{comp2}) we find the wave function in coordinate space as
\begin{eqnarray}\label{foo}
\psi(x)=\sqrt{\frac{\sqrt{\beta}}{\pi}}
\int_{-\frac{\pi}{2\sqrt{\beta}}}^{+\frac{\pi}{2\sqrt{\beta}}}
e^{\frac{i px}{\hbar}}\phi(p)\mathrm{d} p.
\end{eqnarray}

However, since the uncertainties for the eigenfunctions of the
position operator is zero, i.e. $\Delta X_{|x\rangle}=0$,
$|x\rangle$ cannot be the physical solution. So, following Kempf
\emph{et al.} we define the maximal localization states
$|\phi^{\mathrm{ML}}_\xi\rangle$ with the following properties
\cite{12}:
\begin{eqnarray}\label{miangin}
\langle \phi^{\mathrm{ML}}_\xi|X|\phi^{\mathrm{ML}}_\xi\rangle=\xi,
\end{eqnarray}
and
\begin{eqnarray}\label{deltaXML}
\Delta X_{|\phi^{\mathrm{ML}}_\xi\rangle}=(\Delta
X)_{min}=\hbar\sqrt{\beta}.
\end{eqnarray}
These states also satisfy
\begin{eqnarray}
\bigg(X-\langle X\rangle+\frac{\langle[X,P]\rangle}{2(\Delta
P)^2}\left(P-\langle P\rangle\right)\bigg)|\phi\rangle=0,
\end{eqnarray}
where $\langle[X,P]\rangle= i\hbar\left(1+\beta (\Delta
P)^2+\beta\langle P\rangle^2 \right)$. Thus, in the momentum space
the above equation takes the form
\begin{eqnarray}
\left[i\hbar\frac{\partial}{\partial p}-\langle
X\rangle+i\hbar\frac{1+\beta (\Delta P)^2+\beta\langle
P\rangle^2}{2(\Delta
P)^2}\left(\frac{\tan\left(\sqrt{\beta}p\right)}{\sqrt{\beta}}-\langle
P\rangle\right)\right]\phi(p)=0,
\end{eqnarray}
which has the solution
\begin{eqnarray}
\phi(p)= \mathcal{N}\exp\Bigg[\left(-\frac{i}{\hbar}\langle
X\rangle+\frac{1+\beta (\Delta P)^2+\beta\langle
P\rangle^2}{2(\Delta P)^2}\langle P\rangle\right)p
+\left(\frac{1+\beta (\Delta P)^2+\beta\langle P\rangle^2}{2(\Delta
P)^2}\right)
\frac{\ln\left[\cos\left(\sqrt{\beta}p\right)\right]}{\beta}\Bigg].
\end{eqnarray}
To find the absolutely maximal localization states we need to choose
the critical momentum uncertainty $\Delta P=1/\sqrt{\beta}$ that
gives the minimal length uncertainty and take $\langle P\rangle=0$,
i.e.,
\begin{eqnarray}\label{psiML}
\phi^{\mathrm{ML}}_\xi(p)=
\mathcal{N}\cos\left(\sqrt{\beta}p\right)e^{\frac{-i p\xi}{\hbar}},
\end{eqnarray}
where the normalization factor is given by
\begin{eqnarray}
\mathcal{N}=\sqrt{\frac{2\sqrt{\beta}}{\pi}}.
\end{eqnarray}
It is straightforward to check that $\phi^{\mathrm{ML}}_\xi(p)$
exactly satisfies (\ref{miangin}) and (\ref{deltaXML}). Because of
the fuzziness of space, these maximal localization states are not
mutually orthogonal.
\begin{eqnarray}
\langle \phi^{\mathrm{ML}}_{\xi'}|\phi^{\mathrm{ML}}_\xi\rangle=
\mathcal{N}^2\int_{-\frac{\pi}{2\sqrt{\beta}}}^{+\frac{\pi}{2\sqrt{\beta}}}\mathrm{d}
p\cos^2\left(\sqrt{\beta}p\right)e^{\frac{-i
p(\xi-\xi')}{\hbar}}=\frac{8\beta^{3/2}\hbar^3}{\pi}\frac{\sin\left[\frac{\pi(\xi-\xi')}{2\hbar\sqrt{\beta}}\right]}{(\xi-\xi')^3-4\beta\hbar^2(\xi-\xi')}.
\end{eqnarray}
To find the quasiposition wave function $\chi(\xi)$, we  define
\begin{eqnarray}
\chi(\xi)\equiv\langle\phi^{\mathrm{ML}}_\xi|\phi\rangle,
\end{eqnarray}
where in the limit $\beta\rightarrow0$ it goes to the ordinary
position wave function $\chi(\xi)=\langle\xi|\phi\rangle$. Now the
transformation of the wave function in the momentum representation
into its counterpart quasiposition wave function is
\begin{eqnarray}\label{eqpsi}
\chi(\xi)&=&\mathcal{N}\int_{-\frac{\pi}{2\sqrt{\beta}}}^{+\frac{\pi}{2\sqrt{\beta}}}\mathrm{d}
p\cos\left(\sqrt{\beta}p\right)e^{\frac{i
p\xi}{\hbar}}\phi(p), \\
&=&\frac{1}{\sqrt{2}}\left[\psi(\xi+\hbar\sqrt{\beta})+\psi(\xi-\hbar\sqrt{\beta})\right]
.\label{psiF}
\end{eqnarray}
So the quasiposition wave function at $\langle X\rangle=\xi$ is the
superposition of the coordinate space wave functions at
$\xi+\hbar\sqrt{\beta}$ and $\xi-\hbar\sqrt{\beta}$. In other words,
the quasiposition wave function is the result of the interference of
two coordinate space wave functions.

\section{Boundary conditions}
In this section, we discuss how the boundary conditions in
quasiposition space can be determined by fixing the boundary
conditions in coordinate space.

\subsection{Dirichlet boundary condition}
Consider the following Dirichlet boundary condition in coordinate
space
\begin{eqnarray}\label{psi}
\psi(x)\bigg|_{x=\xi_0}=0,
\end{eqnarray}
which gives
\begin{eqnarray}\label{con}
\int_{-\frac{\pi}{2\sqrt{\beta}}}^{+\frac{\pi}{2\sqrt{\beta}}}e^{i\frac{
p}{\hbar}\xi_0}\phi(p)\mathrm{d} p=0.
\end{eqnarray}
Now to first order in the GUP parameter equation (\ref{psiF}) for
continuously differentiable coordinate space wave functions implies
\begin{eqnarray}
\chi(\xi)\bigg|_{\xi=\xi_0}=
\frac{\hbar^2\beta}{\sqrt{2}}\psi''(\xi_0)+{\mathcal O}(\beta^2),
\end{eqnarray}
which fixes the quasiposition wave functions at $\xi=\xi_0$. Note
that, for the following class of the coordinate space solutions
\begin{eqnarray}\label{z0}
\psi(x)= \left\{
  \begin{array}{l}
  A\sin(\omega x)+B\cos(\omega x),   \\
  \sin^n(\omega x),  \\
  \cos^n(\omega x),
  \end{array}\right.
\end{eqnarray}
we exactly have
\begin{eqnarray}
\chi(\xi)\bigg|_{\xi=\xi_0}=\psi(x)\bigg|_{x=\xi_0}=0,
\end{eqnarray}
where $\xi_0$ are the zeros of Eq.~(\ref{z0}). So, for these cases,
the quasiposition wave functions obey the same boundary conditions
as coordinate space wave functions.

\subsection{Neumann boundary condition}
The Neumann  boundary condition determines the values that the
derivative of a wave function is to take on the boundary of the
domain. Let us consider the following boundary condition in
coordinate space
\begin{eqnarray}\label{psi22}
\psi'(x)\bigg|_{x=\xi_0}=0,
\end{eqnarray}
where prime denotes the derivation with respect to the argument.
This equation is equivalent to
\begin{eqnarray}
\int_{-\frac{\pi}{2\sqrt{\beta}}}^{+\frac{\pi}{2\sqrt{\beta}}}e^{i\frac{
p}{\hbar}\xi_0}p\phi(p)\mathrm{d} p=0.
\end{eqnarray}
Also, using Eq.~(\ref{eqpsi}) we have
\begin{eqnarray}
\chi'(\xi)=\frac{1}{\sqrt{2}}\left[\psi'(\xi+\hbar\sqrt{\beta})+\psi'(\xi-\hbar\sqrt{\beta})\right].
\end{eqnarray}
Therefore, to first order in the GUP parameter we obtain
\begin{eqnarray}
\chi'(\xi)\bigg|_{\xi=\xi_0}=
\frac{\hbar^2\beta}{\sqrt{2}}\psi'''(\xi_0)+{\mathcal O}(\beta^2).
\end{eqnarray}
Moreover, for the coordinate space wave functions presented in
Eq.~(\ref{z0}) we exactly find
\begin{eqnarray}
\chi'(\xi)\bigg|_{\xi=\xi_0}=\psi'(x)\bigg|_{x=\xi_0}=0.
\end{eqnarray}
So, both $\chi(\xi)$ and $\psi(x)$ satisfy the same Neumann boundary
condition.

Now let us elaborate the correspondence between the uncertainties in
position and the imposition of localized boundary conditions. In the
GUP framework, it is not possible to measure the position of a
particle more accurate than $(\Delta X)_{min}$. So we cannot define
the potentials with infinitely sharp boundaries. In fact, the
position of these boundaries can be only specified within this
uncertainty. As it is shown in Ref.~\cite{10}, the potentials with
infinitely sharp boundaries such as the particle in a box potential
cannot be properly defined in the GUP framework with respect to
ordinary quantum mechanics.

\section{Conclusions}
In this paper, we have investigated the issue of the boundary
conditions in deformed quantum mechanics which implies a minimal
length uncertainty proportional to the Planck length. We found the
coordinate space wave functions, maximal localization states, and
quasiposition wave functions using a formally self-adjoint
representation. We indicated that the position operator $X$ is
merely symmetric and the momentum operator $P$ is truly self-adjoint
which agrees with Ref.~[13]. The maximal localization states are the
physical states and obey the minimal length uncertainty, i.e.,
$\Delta X_{|\phi^{\mathrm{ML}}_\xi\rangle}=\hbar\sqrt{\beta}$. We
showed that the boundary conditions in coordinate space specify the
boundary conditions in quasiposition space and found the exact
relations for both Dirichlet and Neumann boundary conditions. Also,
for a particular class of solutions, the boundary conditions are
found to be the same in coordinate and quasiposition spaces. In
fact, because of Eq.~(\ref{deltaXML}) the quasiposition wave
functions $\langle\phi^{\mathrm{ML}}_\xi|\phi\rangle$ and their
boundary conditions contain the standard physical interpretation.


\begin{thebibliography}{99}
\bibitem{felder}     S. Hossenfelder, Living Rev. Relativity \textbf{16}, 2 (2013).
\bibitem{1}D. Amati, M. Ciafaloni and G. Veneziano, Phys. Lett. B \textbf{216}, 41 (1989).
\bibitem{2}K. Konishi, G. Paffuti and P. Provero, Phys. Lett. B \textbf{234}, 276 (1990).
\bibitem{3}M. Maggiore, Phys. Lett. B \textbf{304}, 65 (1993).
\bibitem{4}F. Scardigli, Phys. Lett. B \textbf{452}, 39 (1999).
\bibitem{5}M. Maggiore, Phys. Rev. \textbf{49}, 5182 (1994).
\bibitem{6}S. Ghosh and S. Mignemi, Int. J. Theor. Phys. \textbf{50}, 1803 (2011).
\bibitem{7}C. Castro, Found. Phys. Lett. \textbf{10}, 273 (1997).
\bibitem{8}S. Hossenfelder, M. Bleicher, S. Hofmann, J. Ruppert, S. Scherer and H. Stoecker, Phys. Lett. B \textbf{575}, 85 (2003).
\bibitem{9}C. Bambi and F.R. Urban, Class. Quant. Grav. \textbf{25}, 095006 (2008).
\bibitem{10}P. Pedram, Phys. Rev. D \textbf{85}, 024016 (2012).
\bibitem{11}S. Das and E.C. Vagenas, Phys. Rev. Lett. \textbf{101}, 221301 (2008).
\bibitem{12}A. Kempf, G. Mangano, R.B. Mann, Phys. Rev. D \textbf{52}, 1108 (1995).
\bibitem{13}D.J. Gross and P.F. Mende, Nucl. Phys. B \textbf{303}, 407 (1988).
\bibitem{Mead}C.A. Mead, Phys. Rev. \textbf{135}, B849 (1964).
\bibitem{Wigner} H. Salecker and E.P. Wigner, Phys. Rev. \textbf{109}, 571 (1958).
\bibitem{Magg} M. Maggiore, Phys. Lett. B \textbf{304}, 65 (1993).
\bibitem{Padmanabhan1}T. Padmanabhan, Ann. Phys. (N.Y.) \textbf{165}, 38 (1985).
\bibitem{Padmanabhan2}T. Padmanabhan, Gen. Relativ. Gravit. \textbf{17}, 215 (1985).
\bibitem{hooft} G. 't Hooft, and M. Veltman, Ann. Inst. Henri Poincare A \textbf{20}, 69 (1974).
%%%%%%%%%%%%%%%%%%%%%%%%%%%%%%%%%%%%%%%%%%%%%%%%%%%
\bibitem{Adler} R.J. Adler and D.I. Santiago, Mod. Phys. Lett. A \textbf{14}, 1371 (1999).
\bibitem{laser4}     I.~Pikovski, M.R.~Vanner, M.~Aspelmeyer, M.~Kim, C.~Brukner, Nature Physics \textbf{8}, 393 (2012).
\bibitem{laser3}     G.~Amelino-Camelia, J.~Ellis, N.E.~Mavromatos, D.V.~Nanopoulos, and S.~Sarkar, Nature \textbf{393}, 763 (1998).
\bibitem{laser3-2}   U.~Jacob and T.~Piran, Nature Physics \textbf{7}, 8790 (2007).
%%%%%%%%%%%%%%%%%%%%%%%%%%%%%%%%%%%%%%%%%%%%%%%%%%%%

\end{thebibliography}
\end{document}